\documentstyle[preprint,aps]{revtex}
\tightenlines
\input{psfig}
\begin{document}
\preprint{UNDPDK-99-01}
\title{Possible Sources of a Transient Natural Neutrino Flux}
\author{J.M.~LoSecco}
\address{University of Notre Dame, Notre Dame, Indiana 46556}
\date{March 10, 1999}
\maketitle
\begin{abstract}
Recent observations of variability in the ``atmospheric neutrino'' rate place
constraints on possible
explanations of the atmospheric neutrino anomaly.  Many proposed solutions
to the problem are static and can not be adapted to the temporal
dependence observed.  Nonatmospheric sources may be needed to explain the
variations.\\
Subject headings: Cosmic Rays --- Elementary Particles --- Neutrino Oscillations
\\
\end{abstract}


The atmospheric neutrino anomaly is the observation that the rate of
interaction of muon and electron neutrinos in underground detectors do not
match expectations.  The observations indicate that the muon neutrino rate
is below expectations and the electron rate seems to be above expectations.

Observations of the effect seem to have been roughly constant from at least
1984 \cite{imb} to 1997 \cite{superk}.  Recently \cite{dpf} {\em preliminary}
results reported by the Super Kamioka detector are significantly different from
prior measurements, including earlier Super Kamioka
observations \cite{superk,mue}.

While it is generally realized that the new Super Kamioka
results \cite{dpf} differ from
earlier reports \cite{superk,mue} those results are all correlated
since the data sample is cumulative.  All events present in earlier
Super Kamioka analyses have been included in subsequent ones.  The 736
days of reference \cite{dpf} includes the 414.4 days of reference \cite{mue}.
To fully
understand the nature of the variation one needs to isolate independent
data samples.  While there are many ways to do this, one that
maximizes the statistical sensitivity of the comparison is to look
at the event sample collected since the first Super Kamioka publication
of the atmospheric neutrino anomaly \cite{mue}, which was based on 25.5
kt-yrs.  If the \cite{mue} sample is subtracted from the
\cite{dpf} sample one is left with an exposure of 321.6 days for the recent
data.

Table \ref{tab1} lists the sub-GeV events from these three samples, \cite{dpf},
\cite{mue} and the difference.  The last 2 columns of table \ref{tab1}
represent independent observations of atmospheric neutrinos in the same
detector during two non-overlapping time intervals.  The newer sample is
an exposure from November 1997 to November 1998 ($\approx$88\% live time).
Only the statistical error on $R$ is quoted in the table.

The exposure has been listed in days of live time for this 22.5 kiloton
detector.  The rates are all calculated per day of live time.  The smallest
of these samples, the 321.6 day one, is about 19.8 kt-yr which is more than
twice as large as any previous observation, other than Super Kamioka.

Most notable about the earlier and later data samples is the 12$\pm$3\% drop
in overall event rate and the 18$\pm$5\% drop in the ``electron neutrino''
interaction rate.  Misclassification of events as $\mu$ or $e$ or as
single or multi-ring might be attributable to systematic problems with the
detector.  But a significant drop in the overall event rate, for events well
above threshold (this detector sees solar neutrinos) should be taken
seriously.  The 8$\pm$3\% drop in the single ring rate is due to the
3.8 $\sigma$ drop in the $e$ rate.  Table \ref{tab1} also shows 
a 20$\pm$5\% drop in the multi-ring rate.
The muons manifest a not significant 3$\pm$5\% rise.

The drop in electron rate was welcomed by those who favor
a neutrino oscillation interpretation.  Low $\Delta m^{2}$ solutions are not
compatible \cite{prob,RE} with the low value of $R$ (0.61) previously
reported \cite{superk,mue,kam1,imb3}.
But the reason for the rise in $R$, which is a drop in the electron neutrino
rate, is not compatible with the most popular oscillation interpretations
($\nu_{\mu}\rightarrow\nu_{\tau}$ or $\nu_{\mu}\rightarrow\nu_{s}$).
$R$ is a ratio of observed over expected.  References \cite{dpf} and
\cite{mue} also itemize the expected values for the quantities in our
table \ref{tab1}.  The expected portion of $R$ is unchanged for the sub-GeV
data.  The increase in $R$ is mainly due to the drop in the ``electron
neutrino'' interaction rate.

Very few of the suggested solutions to the atmospheric neutrino
problem\cite{what} can support a non static rate.  While the flux of
atmospheric neutrinos is expected to be modulated by the effect of the solar
wind on the Earth's magnetosphere such effects happen slowly, with the sunspot
cycle.  A change in the overall cosmic ray flux impacting the atmosphere is
not expected to have an influence on the $\mu/e$ ratio in the neutrino flux.
The atmospheric $\mu/e$ ratio is determined by static local factors such as
the pion and muon lifetime and the altitude at which production takes place.

For comparison purposes
table \ref{tab2} summarizes the Super Kamioka multi-GeV data samples from
the same three time
periods.  The multi-GeV sample is the event sample above 1.3 GeV.
It is noteworthy that there is no statistically significant
variation in the multi-GeV data.  The multi-GeV electron sample has not
declined.  The value of $R$ seems to be static in this energy range.

Common terrestrial sources of neutrinos can not account for the changes.
Nuclear reactors produce electron antineutrinos with energies below 8 MeV
which would never get into the sample illustrated in table \ref{tab1}.
Accelerators would produce an obviously directional source, most of which
would be of the muon neutrino type.  But the muon rate seems to be unchanged.

Possibilities considered below are, neutrino production off of a diffuse
cloud, a temporally compact $\nu_{e}$ burst or fluctuations in some
ambient dark matter which has a signature similar to electron neutrino
interactions.
Variability of the ``atmospheric neutrino'' rate and composition is a strong
indicator that at least some of the signal is not from neutrinos of
atmospheric origin.  The lack of significant variability in the muon rate
is difficult to understand.  One could expect the electron neutrino to
muon neutrino rate from a possible cosmic source to be different from the
value of approximately 0.5 expected from atmospheric pion and muon decay.

Tenuous astrophysical clouds could be a source of neutrinos with a significant
neutron decay contribution which would increase the electron neutrino
fraction.  But such a source would also produce comparable
numbers of muon neutrinos.  If the earth emerged from a diffuse cloud that was
producing high energy neutrinos off of the ambient cosmic ray flux the
observed rates of both electron and muon neutrinos should have dropped, but
not by the same amount.
Such a diffuse source would produce a flux similar to the atmospheric flux
augmented by neutrinos from neutron decay and production by cosmic rays
which are too low in energy to penetrate the Earth's magnetic field.
But the observations are consistent with no change in the muon neutrino flux
which argues against this hypothesis.

A diffuse cloud source would enhance all neutrino fluxes that are path length
limited in the atmosphere.  In addition to neutron decays, one would expect
neutrinos coming from high energy pions and muons that strike the ground
(or sea) on Earth, to be greatly enhanced.  The absence of any
significant change in the multi-GeV sample, table \ref{tab2} makes this
hypothesis unlikely.

A temporally compact $\nu_{e}$ pulse of several years duration could also
account for the observations but there is no known source of such
an astrophysical structure.  Supernovae produce such pulses with much lower
energy neutrinos, 10's of MeV and with a duration of seconds not years.
By tailoring the energy spectrum such a pulse could be made compatible with
the drop in multi-ring events.

Dark matter had been ruled out as the source of the excess ``electron
neutrino'' signal\cite{dark} since the energy density required would be too
high.  (By dark matter we mean an ambient weakly interacting species that
either enters the detector and reacts in a manner producing a signal similar to
electron neutrino interactions or such a particle that decays in the detector
producing a signal similar to electron neutrino interactions.)
But if prior observations were attributable to local
fluctuations in the dark matter density this limit could be evaded.
If the previously reported high value of the ``dark matter'' density estimated
from the atmospheric neutrino anomaly was not indicative of the global average,
the closure argument can be avoided.

Interpretation of the ``atmospheric neutrino'' anomaly as the observation of
a transient nonatmospheric source eliminates one source of confusion.
Examination of the differences in energy distributions and directional
distributions may ascertain if the excess events were neutrino
induced or were due to a more exotic interaction.  Differences between the
multi-ring and single ring rates, as shown in table \ref{tab1} may also
be associated with either an energy spectrum difference or a new form of
interaction.  The fact that a comparable drop has been seen in electron like
and multi-ring events may be of some significance.

The value of $R$ reported in the newest sub-GeV sample is low,
0.76$\pm$0.04(stat.)$\pm$0.06 (syst.) so
a second source for part of the anomaly may still be present.  But since the
variability has just been noted it would be prudent to wait until a
static value for the $\mu/e$ ratio is again established before one is
motivated to search for a second solution.  Comparison of the intermediate
sub-GeV neutrino sample of 33 kt-yrs \cite{superk} with the 25.5 \cite{mue} and
45.4 kt-yr \cite{dpf} samples used in this paper indicates that the flux
rates were varying during the one year time interval illustrated in the
third column of our tables.

\section*{Acknowledgements}
I would like to thank M.~Messier for making his transparencies of the DPF talk
available to me.  D.~Casper, J.~Learned and M.~Messier have been helpful in
dispelling my reservations about systematic errors and assuring me that
errors in the live time are negligible.  I thank M.~Weinstein
for some discussions about possible transient celestial neutrino sources.
I.~Bigi and P.~Harrison have been a source of inspiration and encouragement.
I thank J.~Poirier for a careful reading of the manuscript.

\begin{table}
\begin{tabular}{lrrr}
& DPF\cite{dpf} & Phys. Lett.\cite{mue} & New Sample \\ \hline
Single Ring & 3224 & 1883 & 1341 \\
e-like & 1607 & 983 & 624 \\
$\mu$-like & 1617 & 900 & 717 \\
Multi-Ring & 1271 & 784 & 487 \\
Total & 4495 & 2667 & 1828 \\
$\mu/e$ & 1.01$\pm$0.04 & 0.92$\pm$0.04 & 1.15$\pm$0.06 \\
$R=(\mu/e)_{Obs}/(\mu/e)_{MC}$ & 0.67$\pm$0.02 & 0.61$\pm$0.03
& 0.76$\pm$0.04 \\
Exposure (days) & 736 & 414.4 & 321.6 \\
Event Rate & 6.11$\pm$0.09 & 6.44$\pm$0.12 & 5.68$\pm$0.13 \\
Single Ring Rate & 4.38$\pm$0.08 & 4.54$\pm$0.10 & 4.17$\pm$0.11 \\
e rate & 2.18$\pm$0.05 & 2.37$\pm$0.08 & 1.94$\pm$0.08 \\
$\mu$ rate & 2.20$\pm$0.05 & 2.17$\pm$0.07 & 2.23$\pm$0.08 \\
Multi-Ring Rate & 1.73$\pm$0.05 & 1.89$\pm$0.07 & 1.51$\pm$0.07 \\
\end{tabular}
\caption{\label{tab1}Comparison of Super Kamioka Sub-GeV Samples}
\end{table}

\begin{table}
\begin{tabular}{lrrr}
& DPF\cite{dpf} & Phys. Lett.\cite{mult} & New Sample \\ \hline
Single Ring & 687 & 394 & 293 \\
e-like &  386 & 218 & 168 \\
$\mu$-like & 301 & 176 & 125 \\
Multi-Ring & 737 & 398 & 339 \\
Total & 1424 & 792 & 632 \\
Partially Contained (PC) & 374\cite{red} & 230 & 144\cite{red} \\
$\mu/e$ & 0.78$\pm$0.06 & 0.81$\pm$0.08 & 0.74$\pm$0.09 \\
$R=(\mu/e)_{Obs}/(\mu/e)_{MC}$ \cite{nopc} & 0.65$\pm$0.05 & 0.64$\pm$0.07 &
0.66$\pm$0.08 \\
Exposure (days) & 736 & 414.4 & 321.6 \\
Event Rate & 1.93$\pm$0.05 & 1.91$\pm$0.07 & 1.97$\pm$0.08 \\
Single Ring Rate & 0.93$\pm$0.04 & 0.95$\pm$0.05 & 0.91$\pm$0.05 \\
e rate & 0.52$\pm$0.03 & 0.53$\pm$0.04 & 0.52$\pm$0.04 \\
$\mu$ rate & 0.41$\pm$0.02 & 0.42$\pm$0.03 & 0.39$\pm$0.03 \\
Multi-Ring Rate & 1.00$\pm$0.04 & 0.96$\pm$0.05 & 1.05$\pm$0.06 \\
PC Rate & 0.55$\pm$0.03 & 0.56$\pm$0.04 & 0.53$\pm$0.04 \\
\end{tabular}
\caption{\label{tab2}Comparison of Super Kamioka Multi-GeV Samples}
\end{table}


\begin{references}
\bibitem{imb} T.J.~Haines {\em et al.}, Phys.~Rev.~Lett. {\bf 57}, 1986
(1986)\\
E.L.~Shumard, ``Search for Nucleon Decay $p \rightarrow
\nu K^{+}$, $p \rightarrow \nu K^{*+}$ and $n \rightarrow \nu K^{*0}$'',
Ph.D. thesis, The University of Michigan (1984).
\bibitem{superk}  Y.~Fukuda {\em et al.},
Phys.~Rev.~Lett. {\bf 81}, 1562 (1998).
\bibitem{dpf} M.~Messier, Talk presented at the 1999 DPF meeting, January 1999.\\
http://hep.bu.edu/$\sim$messier/dpf/index.html
\bibitem{mue} Y.~Fukuda {\em et al.},
Phys.~Lett. {\bf B433}, 9 (1998).
\bibitem{prob} J.M.~LoSecco, ``Problems with Atmospheric Neutrino
Oscillations'',  hep-ph/9807359 (to be published).
\bibitem{RE} J.M.~LoSecco, ``$\Delta m^{2}$ Limits from $R(E_{\nu})$'',
hep-ph/9807432 (to be published). 
\bibitem{kam1}
K.S.~Hirata {\em et al.}, Phys.~Lett. {\bf B205}, 416 (1988)\\
K.S.~Hirata {\em et al.}, Phys.~Lett. {\bf B280}, 146 (1992).
\bibitem{imb3}
D.~Casper {\em et al.}, Phys.~Rev.~Lett. {\bf 66}, 2561 (1991).\\
R.~Becker-Szendy {\em et al.}, Phys.~Rev. {\bf D46}, 3720 (1992).
\bibitem{what} J.M.~LoSecco, ``What the Atmospheric Neutrino Anomaly is Not'',
hep-ph/9809499 (to be published).
\bibitem{dark} J.~LoSecco,
Phys.~Rev. {\bf D56}, 4416 (1997).
\bibitem{mult}  Y.~Fukuda {\em et al.},
Phys.~Lett. {\bf B436}, 3 (1998).
\bibitem{red} The DPF PC sample is based on an exposure of 685 days so the
difference sample for the PC events is based on 270.6 days.
\bibitem{nopc} We do not include the PC events in our calculation of $R$.
\end{references}
\end{document}